\documentclass[twocolumn,showpacs,preprintnumbers,amsmath,amssymb]{revtex4}

\usepackage{graphicx}
\usepackage{dcolumn}
\usepackage{bm}

\newcommand{\EQ}{\begin{equation}}
\newcommand{\EN}{\end{equation}}
\newcommand{\EQA}{\begin{eqnarray}}
\newcommand{\ENA}{\end{eqnarray}}

\newcommand{\Eq}[1]{Eq.~(\ref{#1})}
\newcommand{\Eqs}[2]{Eqs~(\ref{#1}) and~(\ref{#2})}
\newcommand{\eqs}[2]{(\ref{#1}) and~(\ref{#2})}

\newcommand{\Fig}[1]{Fig.~\ref{#1}}

\newcommand{\Tab}[1]{Table~\ref{#1}}

\newcommand{\bra}[1]{\langle #1\rangle}
\newcommand{\bbra}[1]{\left\langle #1\right\rangle}

\newcommand{\meanAA}{\overline{\mbox{\boldmath $A$}}}
\newcommand{\meanBB}{\overline{\mbox{\boldmath $B$}}}

\newcommand{\xx}{\mbox{\boldmath $x$} {}}

\newcommand{\uu}{\mbox{\boldmath $u$} {}}

\newcommand{\bb}{\mbox{\boldmath $b$} {}}

\newcommand{\BB}{\mbox{\boldmath $B$} {}}

\newcommand{\AAA}{\mbox{\boldmath $A$} {}}
\newcommand{\aaa}{\mbox{\boldmath $a$} {}}

\newcommand{\JJ}{\mbox{\boldmath $J$} {}}

\newcommand{\ff}{\mbox{\boldmath $f$} {}}

\newcommand{\FF}{\mbox{\boldmath $F$} {}}

\newcommand{\nab}{\mbox{\boldmath $\nabla$} {}}

\newcommand{\DD}{{\rm D} \, {}}
\newcommand{\dd}{{\rm d} {}}

\def\onethird{{\textstyle{1\over3}}}

\newcommand{\ynat}[3]{, Nature {\bf #2}, #3 (#1).}

\newcommand{\yjfm}[3]{, J. Fluid Mech. {\bf #2}, #3 (#1).}

\newcommand{\yprl}[3]{, Phys.\ Rev.\ Lett.\ {\bf #2}, #3 (#1).}

\newcommand{\yjcp}[3]{, J. Comp. Phys. {\bf #2}, #3 (#1).}

\newcommand{\ygrl}[3]{, Geophys. Res. Lett. {\bf #2}, #3 (#1).}

\newcommand{\yapj}[3]{, Astrophys. J. {\bf #2}, #3 (#1).}

\newcommand{\ypp}[3]{, Phys. Plasmas {\bf #2}, #3 (#1).}

\newcommand{\ypf}[3]{, Phys. Fluid {\bf #2}, #3 (#1).}
\newcommand{\yphy}[3]{, Physica {\bf #2}, #3 (#1).}
\newcommand{\ygafd}[3]{, Geophys. Astrophys. Fluid Dynam. {\bf #2}, #3 (#1).}

\newcommand{\yjour}[4]{, #2 {\bf #3}, #4 (#1).}

\newcommand{\yprep}[2]{, (preprint).}

\begin{document}

\preprint{NORDITA 2001-30 AP}

\title{The effect of hyperdiffusivity on turbulent dynamos with helicity}
\author{Axel Brandenburg}
\affiliation{NORDITA, Blegdamsvej 17, DK-2100 Copenhagen \O, Denmark}
\author{Graeme R. Sarson}
\affiliation{Department of Mathematics, University of Newcastle upon Tyne, NE1 7RU, UK}
\date{\today}

\begin{abstract}
In numerical studies of turbulence, hyperviscosity
is often used as a tool to extend
the inertial subrange and to reduce the dissipative subrange.
By analogy, hyperdiffusivity (or hyperresistivity) is sometimes
used in magnetohydrodynamics. The underlying assumption is that
only the small scales are affected by this manipulation.
In the present
paper, possible side effects on the evolution of the large scale magnetic field
are investigated.  It is found that for turbulent flows with helicity,
hyperdiffusivity causes the dynamo-generated magnetic field to saturate
at a higher level than normal diffusivity.
This result is successfully interpreted in terms of magnetic helicity
conservation, which also predicts that
full saturation is only reached after a time comparable to
the large scale magnetic (hyper)diffusion time.
\end{abstract}

\maketitle

In theoretical studies of Navier-Stokes turbulence the ordinary viscosity
operator, $\nu\nabla^2\bm{u}$, is sometimes replaced by
$(-1)^{n-1}\nu_n\nabla^{2n}\uu$, where $\nu_n$ is a
hyperviscosity of order $n$.
The use of hyperviscosity
has the advantage of making the transition from the inertial
subrange to the viscous subrange shorter \cite{BLSB81}. 
At the same time, however, it has the notorious disadvantage of tempering
possibly major parts of the inertial subrange. An example is the so-called
bottleneck effect that leads to significantly shallower power spectra
at high wavenumbers near and before the viscous subrange
\cite{Fal94}.  
When early simulations using the piecewise parabolic method
showed such bottleneck effect \cite{PPW92}, it was unclear whether
this effect was real or just a consequence of hyperviscosity.
In the context of magnetohydrodynamics,
recent comparisons between direct and hyperviscous simulations
point now to the latter possibility \cite{BM00}.

The use of hyperviscosity is indeed quite popular in studies of hydromagnetic
turbulence where, in addition to viscosity, the ordinary magnetic diffusivity is
replaced by hyperdiffusivity \cite{BM00,BSC98,BS01,MFP81}.
A strong artificial bottleneck effect occurs when hyperdiffusivity is used \cite{BM00}.
This is particularly clear in two dimensions at very high resolution
\cite{BSC98}, although a weak bottleneck effect occurs even without
any hyperdiffusive effects \cite{BS01}.
The use of hyperviscosity and hyperdiffusivity has also led to
significant controversy \cite{ZJ97} 
in models of the Earth's dynamo \cite{GR95}. 
At the center of the controversy is the effect of hyperviscosity on
the asymptotic behavior at small Ekman number (low viscosity).  This
can also affect conclusions regarding the relative importance of the
Lorentz force, and the relevance of Taylor's constraint,
both matters of great importance for geodynamo theory.
There are also examples where the use of hyperdiffusivity has moved dynamos
from an $\alpha^2$-regime toward an $\alpha\Omega$-regime \cite{GBT00}.

Generally speaking, hyperviscosity and hyperdiffusivity can lead to rather
ill-understood behavior that tends to diminish its potential advantages.
It is therefore important to clarify exactly how hydromagnetic dynamos are
affected by this approach.
Here, we concentrate on the effects of (magnetic) hyperdiffusivity.
This is done in the context of MHD turbulence;
future work will look at this effect on geodynamo models.

In the present paper we show that, if the fluid motions are helical,
hyperdiffusivity can lead to artificially enhanced saturation amplitudes of the
nonlinear dynamo.
It is at first glance somewhat counterintuitive that the
properties of the {\it large scale} field should depend on the details
of the diffusion operator, which is supposed to affect only the small scales.
In recent years, however, there has been mounting evidence that
large scale dynamos, which usually involve helicity, do depend on the
microscopic diffusion \cite{VC92}. 
This property 
is related to magnetic helicity conservation, which permits magnetic
helicity to change only on a resistive time scale;
see Ref.~\cite{B2001}, hereafter referred to as B2001. 
These processes can be seriously affected by the use of hyperdiffusivity.
Obtaining a detailed understanding of the associated artifacts is crucial
before hyperdiffusivity can be taken as a useful tool in dynamo simulations.
We emphasize that the sensitivity to the use of hyperdiffusion reported in
the present paper is peculiar to {\it large scale} dynamos and does not
apply to small scale dynamos.

The magnetic field evolution is governed by the induction equation,
\begin{equation}
\frac{\partial\BB}{\partial t}=\nab\times(\uu\times\BB) + (-1)^{n-1}\eta_n\nabla^{2n}\BB,
\label{induction}
\end{equation}
with $\nab\cdot\BB=0$, $\eta_n={\rm const}$, and $n=1$ for ordinary
magnetic diffusivity. This equation has to be integrated together with
the momentum and continuity equations which are, for an isothermal
compressible gas with constant speed of sound, $c_{\rm s}$,
\begin{equation}
{\DD\uu\over\DD t}=-c_{\rm s}^2\nab\ln\rho+{\JJ\times\BB\over\rho}
+\FF_{\rm visc}+\ff,
\end{equation}
\begin{equation}
{\DD\ln\rho\over\DD t}=-\nab\cdot\uu,
\end{equation}
where ${\DD/\DD t}=\partial/\partial t+\uu\cdot\nab$ is the
advective derivative, 
$\FF_{\rm visc}={\mu\over\rho}(\nabla^2\uu+\onethird\nab\nab\cdot\uu)$
is the viscous force, $\mu={\rm const}$ is the dynamical viscosity, $c_{\rm s}$
is the isothermal sound speed, $\JJ=\nab\times\BB/\mu_0$ is
the current density, and $\mu_0$ is the vacuum permeability.
As in B2001, the forcing function $\ff(\xx,t)$ is a randomly chosen polarized wave
taken from a band of wavenumbers around the forcing wavenumber $k_{\rm f}$.
The direction and phase of $\ff$ change randomly at each time step.
We solve the equations in a periodic domain of size $L^3$, where
$L=2\pi$.
We adopt nondimensional units
by measuring density in units of the average value, $\rho_0=\bra{\rho}$, length in
units of $1/k_1$, where $k_1=2\pi/L=1$ is the smallest wavenumber,
velocity in units of $c_{\rm s}$, magnetic field in units of
$c_{\rm s}\sqrt{\rho_0\mu_0}$, and time in units of 
$(c_{\rm s} k_1)^{-1}$. 

For numerical solutions, 
the smallest possible value of $\eta_n$ is given by the condition
that the mesh magnetic Reynolds number, based on the smallest scales resolved,
\begin{equation}
R_{\rm m}^{\rm(mesh)}=u_{\rm rms}/(\eta_n k_{\rm Ny}^{2n-1})
\end{equation}
is of order unity. Here, $k_{\rm Ny}=\pi N/L$ is the Nyquist wavenumber
of a mesh with $N$ points.

\begin{figure}\includegraphics[width=.5\textwidth]{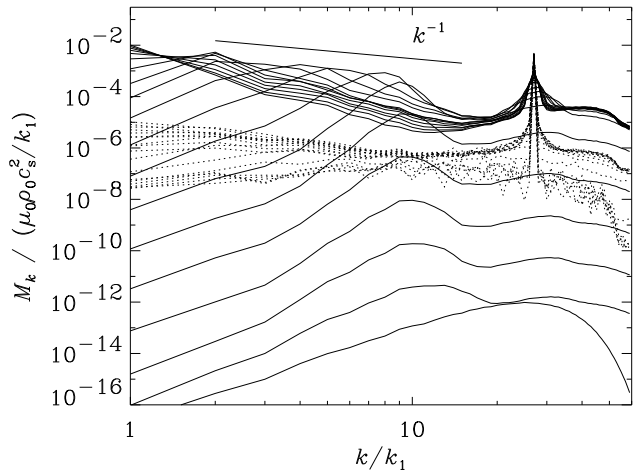}\caption[]{
Evolution of magnetic energy spectra in Run~A (with ordinary
magnetic diffusion) in equidistant time intervals
at $t=0$ (lowest curve), $t=80$ (next one higher up), until $t=1200$
(peaking at the very top left). The dotted lines give the kinetic energy.
}\label{Fpspec_allo}\end{figure}

As a reference model, 
we consider a calculation with ordinary magnetic diffusivity,
$\eta_1=10^{-4}$ and a forcing wavenumber $k_{\rm f}=27$ (Run~A).
We adopt a dynamical viscosity of $\mu=10^{-2}$, so the magnetic Prandtl number
is 100. In \Fig{Fpspec_allo} we show spectra at different times.
Consistent with earlier results (B2001), magnetic energy grows owing
to dynamo action with spectral peaks at the forcing scale, $k=k_{\rm f}$,
and at some intermediate scale, $k\approx 9$.
By $t \approx 480$ (seventh curve from the bottom of \Fig{Fpspec_allo})
the intermediate scale field has reached equipartition with the kinetic
energy;  the field continues to grow, however, and now evolves towards
larger scales under a $k^{-1}$ envelope.

\begin{figure}\includegraphics[width=.5\textwidth]{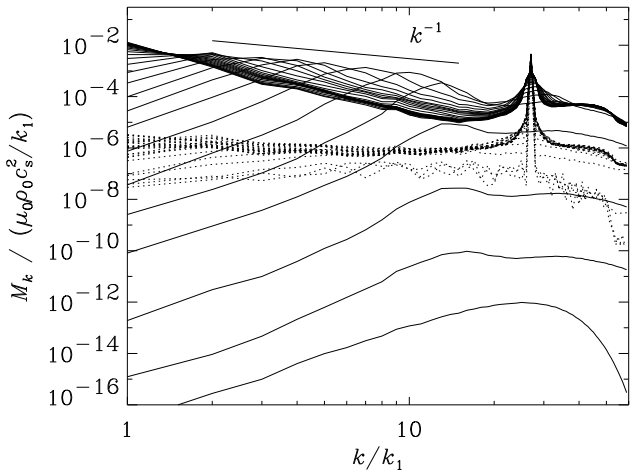}\caption[]{
Evolution of magnetic energy spectra in the hyperdiffusive Run~B
in equidistant time intervals
at $t=0$ (lowest curve), $t=40$ (next one higher up), until $t=2080$
(peaking at the very top left). The dotted lines give the kinetic energy.
}\label{Fpspec_allo_Hyp3b}\end{figure}

Qualitatively similar behavior is found for the hyperdiffusive Run~B, with
$\eta_2=3\times10^{-8}$; see \Fig{Fpspec_allo_Hyp3b}.  This case,
which has similar diffusion at the Nyquist wavenumber to Run~A, also
exhibits a secondary peak at some intermediate wavenumber.
Following B2001, cf.\ Eq.~(36)--(39) therein, we identify this with the
wavenumber where growth rate of a corresponding $\alpha^2$ dynamo model
is maximum.
In the initial
kinematic stage, the position of the secondary peak is approximately
constant in time ($k_{\max}\approx9$ for Run~A and $k_{\max}\approx14$
for Run~B). When the magnetic energy reaches equipartition with the kinetic
energy, $E_{\rm kin}$, the secondary peak begins to travel toward smaller
$k$. A reasonable fit to this migration is given by 
\begin{equation}
k_{\max}^{-1}=\alpha_{\rm trav}(t-t_{\rm sat}),
\end{equation}
where the parameter $\alpha_{\rm trav}$ characterizes the speed at which
the intermediate peak travels. 
The values obtained from fits to these runs
are listed in \Tab{Tsum}, together with some other parameters defined below.
Note that for Run~C, with a lower hyperdiffusivity $\eta_{2}=10^{-8}$,
the speed of the peak has decreased even further. This suggests that
{\it even at intermediate scales} (i.e.\ $k$ less than the initial value of
$k_{\rm max}$), the dynamo process is resistively limited.

\begin{table}[htb]\caption{
Summary of the runs discussed in this paper.
Note that $\alpha_{\rm trav}$ decreases with decreasing values of $\eta_n$.
$\lambda$ is the kinematic growth rate of the magnetic energy,
$H$ and $M$ are, respectively, magnetic helicity and energy during
the kinematic stage, and the parameter $\ell_{\rm skin}$ (defined below)
gives an approximate upper bound for $\ell_{\rm H}\equiv|H|/(2\mu_0M)$.
}\begin{tabular}{lccccccccc}
Run &   $N$   &$n$& $\eta_n$    & $\lambda$ & $k_{\rm f}$ & $\alpha_{\rm trav}$
                                                &  $\ell_{\rm H}$ & $\ell_{\rm skin}$\\
\hline
A   & $120^3$ & 1 &  $10^{-4}$     &0.047  & 27 & $1.1\times10^{-3}$ & 0.035& 0.065 \\ 
B   & $120^3$ & 2 &$3\times10^{-8}$& 0.070 & 27 & $7.3\times10^{-4}$ & 0.018& 0.025 \\ 
C   & $120^3$ & 2 &    $10^{-8}$   & 0.082 & 27 & $3.6\times10^{-4}$ & 0.005& 0.013 \\ 
D   &  $30^3$ & 2 &  $10^{-4}$     &0.078  &  3 &       --           & 0.08 & 0.15  \\ 
\label{Tsum}\end{tabular}\end{table}

\begin{figure}\includegraphics[width=.5\textwidth]{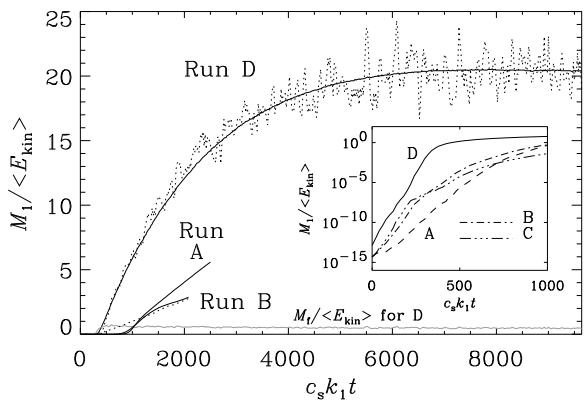}\caption[]{
Evolution of large scale magnetic energy, normalized by the kinetic energy,
for Runs~A, B and D. The dotted lines give the result expected from
\Eq{SaturationLimit2}, as is explained below; these lines show
strong fluctuations because we have used here the actual value of
$M_{\rm f}$. For Run~D, $M_{\rm f}$ is also shown (grey line
at the very bottom). The inset shows the early kinematic
evolution for Runs~A-D in a semi-logarithmic representation.
}\label{Fppm_sat_comp}\end{figure}

Since it is difficult to evolve a simulation at this resolution
over a full large scale diffusion time, $\sim(\eta_n k_1^{2n})^{-1}$,
we now compare with a low resolution run with only $30^3$ mesh points,
$\eta_{n} = 10^{-4}$ (Run~D).
The parameters of this run are chosen so that the factor $k_{\rm f}^{2n-1}$,
which appears in the theory below, is consistent with Run~A.
The large scale magnetic field shows a very prolonged
saturation phase after the saturation of the small scale field
(and the equipartition of magnetic and kinetic energy),
finally equilibrating only after
approximately one large scale diffusion time; see \Fig{Fppm_sat_comp}.

As in B2001, we can interpret the slow saturation behavior in terms of the
magnetic helicity equation. We begin with the uncurled induction equation,
\begin{equation}
{\partial\AAA\over\partial t}=\uu\times\BB
+(-1)^{n-1}\eta_n\nabla^{2n}\AAA-\nab\phi,
\label{inductionA}
\end{equation}
where $\AAA$ is the magnetic vector potential, with $\BB=\nab\times\AAA$,
and $\phi$ is the electrostatic potential which can be chosen arbitrarily
without affecting $\BB$.
For a periodic domain of volume $V$, the magnetic helicity,
\begin{equation}
H=\int_V\AAA\cdot\BB\,\dd V\equiv\bra{\AAA\cdot\BB}\,V,
\end{equation}
where angular brackets denote volume averages, is independent of the
choice of $\phi$. (For the simulations we take $\phi=0$.)
Dotting \Eq{induction} with $\AAA$, and \Eq{inductionA}
with $\BB$, adding the two and averaging, yields
\begin{equation}
{\dd\over\dd t}\bra{\AAA\cdot\BB}=(-1)^n\,
2\eta_n\bbra{(\nabla^{2n}\AAA)\cdot\BB}.
\label{dABn}
\end{equation}
Surface terms are absent because of periodic boundaries.
For this reason, and because $\nab\cdot\BB=0$, the
right hand side of \Eq{dABn} becomes $-2\eta_1\bra{\JJ\cdot\BB}$ when $n=1$.

We now proceed analogously to B2001. Firstly, in the
steady state, $\bra{\AAA\cdot\BB}$ must be constant and so
$\bbra{(\nabla^{2n}\AAA)\cdot\BB}$ must vanish. This happens in such
a way that there are contributions from the forcing scale and the large
scale such that the two terms have opposite sign and cancel.
We calculate the large scale field, $\meanBB=\nab\times\meanAA$, by Fourier
filtering around $k=1$ (using integer bins), 
and the small scale field, $\bb=\nab\times\aaa$,
as $\bb=\BB-\meanBB$.
The characteristic wavenumbers of these scales are $k_{\rm f}$
(for forcing or fluctuating scale) and $k_{\rm 1}$ (for smallest wave number).
Making use of the fact that the magnetic field is nearly fully helical at
small and large scales, we have
\begin{equation}
(-1)^n\bbra{(\nabla^{2n}\aaa)\cdot\bb}\approx\pm k_{\rm f}^{2n-1}\bra{\bb^2},
\label{fulhelf}
\end{equation}
\begin{equation}
(-1)^n\bbra{(\nabla^{2n}\meanAA)\cdot\meanBB}\approx\mp k_{\rm 1}^{2n-1}\bra{\meanBB^2},
\label{fulhelm}
\end{equation}
where the upper and lower signs are for positive and negative signs
of the kinetic helicity of the forcing, respectively.
Thus, the ratio of large to small scale magnetic energies, i.e.\
the degree of superequipartition, is
\begin{equation}
{M_{\rm 1}\over M_{\rm f}}\equiv{\bra{\meanBB^2}\over\bra{\bb^2}}
=\left({k_{\rm f}\over k_{\rm 1}}\right)^{2n-1}>1.
\end{equation}
For $k_{\rm f}=3$, normal diffusion ($n=1$) gives
superequipartition by a factor of 3;
see B2001.
For Run~D ($n=2$) we should have superequipartition by a factor of 27. The
numerical result (\Fig{Fppm_sat_comp},
where $M_{\rm f}/\bra{E_{\rm kin}}\approx0.5$)
gives $M_1/M_{\rm f}\approx44$. As we explain below,
this has to do with the fact that the estimates in
\eqs{fulhelf}{fulhelm} are not quite accurate.
Nevertheless, the effect of hyperdiffusivity on the level 
of superequipartition is clear.

Analogously to B2001 we can also calculate the asymptotic saturation
behavior by using \Eqs{fulhelf}{fulhelm} and assuming
equipartition at small scales, $\bra{\bb^2}\approx\bra{\rho\uu^2}$,
which is expected to hold after the time $t_{\rm s}$ when the
small scale field has saturated.
(The fact that these solutions satisfy 
$M_{\rm f}/\bra{E_{\rm kin}}\approx0.5$, 
rather than strict equipartition,
does not affect the following.)
This gives
\begin{equation}
k_{\rm 1}^{-1}{\dd\over\dd t}\bra{\meanBB^2}=
-2\eta_n k_{\rm 1}^{2n-1}\bra{\meanBB^2}
+2\eta_n k_{\rm f}^{2n-1}\bra{\bb^2},
\end{equation}
which has the solution
\begin{equation}
\bra{\meanBB^2}=\bra{\bb^2}
\left({k_{\rm f}\over k_{\rm 1}}\right)^{2n-1}
\left[1-e^{-2\eta_n k_{\rm 1}^{2n}(t-t_{\rm s})}\right].
\label{SaturationLimit1}
\end{equation}
In the early saturation phase, we have
\begin{equation}
\bra{\meanBB^2}/\bra{\bb^2}\approx 2\eta_n k_{\rm 1}k_{\rm f}^{2n-1}(t-t_{\rm s}).
\end{equation}
Thus, 
Run~A ($\eta_1=10^{-4}$ and $k_{\rm f}=27$) and
Run~D ($\eta_2=10^{-4}$ and $k_{\rm f}=3$) should exhibit 
the same saturation behavior;
this can be approximately verified from the early 
saturation behavior visible in \Fig{Fppm_sat_comp}.

The magnetic field in Run~D actually saturates somewhat
faster than suggested by \Eq{SaturationLimit1}. Again, this is
explained by the observation that the estimates for
the effective values of the wavenumbers in \eqs{fulhelf}{fulhelm}
are not accurate. Good agreement can be achieved
if, instead, we use \Eqs{fulhelf}{fulhelm}
to calculate {\it effective} wavenumbers,
$k_{\rm 1}\rightarrow k_{\rm 1,eff}$ and $k_{\rm f}\rightarrow k_{\rm f,eff}$,
for the large and small scale fields, respectively, and if
we use the actual values for the small scale magnetic energy,
$M_{\rm f}$ (which show a long-term trend, but is also fluctuating).
For Run~D we find $k_{\rm 1,eff}=1.3$ and $k_{\rm f,eff}=4.6$.
Such an enhancement results from hyperdiffusivity which increases
the relative contributions from higher harmonics.
The modified version of \Eq{SaturationLimit1} is then
\begin{equation}
M_{\rm 1}=M_{\rm f}
\left({k_{\rm f,eff}\over k_{\rm 1,eff}}\right)^{2n-1}
\left[1-e^{-2\eta_n k_1 k_{\rm 1,eff}^{2n-1}(t-t_{\rm s})}\right].
\label{SaturationLimit2}
\end{equation}
The evolution predicted by \Eq{SaturationLimit2} is shown
as dotted lines in \Fig{Fppm_sat_comp}.
Note that the time taken for saturation is dependent upon the 
large scale hyperdiffusion time $(\eta_{n} k_{\rm 1, eff}^{2n})^{-1}$,
but that since the large scale field is of approximately unit wavenumber,
$k_{\rm 1,eff} \approx 1$, the hyperdiffusivity has very little effect,
decreasing this time only slightly (cf.\ the true large scale diffusion 
time for this value of $\eta$).
In this respect, hyperdiffusivity is behaving exactly as we would wish;
allowing us to attain low $\eta$ at lesser computational expense,
and with little effect on the physical behavior.

We note that during the {\it kinematic} phase the dynamo
is still growing on a fast dynamical time scale. 
At this stage, the net magnetic helicity remains close to zero,
as it must for the high magnetic Reynolds numbers under consideration.
Berger's inequality \cite{Ber1984} gives an upper limit for the growth
of magnetic helicity, derived by 
bounding the right hand side of \Eq{dABn} via the square root
of Joule dissipation and magnetic energy.
In the presence of hyperdiffusivity this inequality is
\begin{equation}
\ell_{\rm H}\equiv|H|/(2\mu_0M)\le a\ell_{\rm skin},
\label{HMconst}
\end{equation}
where
$\ell_{\rm skin}=(2\eta_n k_{\rm f}^{2n-2}/\lambda)^{1/2}$
is a modified skin depth, $\lambda$ is the kinematic growth rate
of the magnetic energy, and $a$ is a coefficient of order unity.
 From \Tab{Tsum} we see that this constraint is indeed well satisfied
during the kinematic growth phase.

The present results have demonstrated that hyperdiffusivity can have
profound effects on dynamos with helicity. The modifying effects are
well understood, which makes the use of hyperdiffusivity an efficient
tool for numerical studies.
This has allowed us here to show that helical dynamos saturate resistively
both on large and intermediate scales, but not on small scales.

Use of the PPARC supported supercomputers in St Andrews
and Leicester (UKAFF) is acknowledged.


\end{document}